\documentclass[runningheads]{llncs}
\usepackage{graphicx}

\usepackage{xcolor}

\newcommand{\done}[1]{}  
\newcommand{\removeText}[1]{}

\newcommand{\change}[1]{\textcolor{black}{#1}}

\usepackage{booktabs}
\usepackage{array}
\newcolumntype{L}[1]{>{\raggedright\let\newline\\\arraybackslash\hspace{0pt}}m{#1}}
\newcolumntype{C}[1]{>{\centering\let\newline\\\arraybackslash\hspace{0pt}}m{#1}}
\newcolumntype{R}[1]{>{\raggedleft\let\newline\\\arraybackslash\hspace{0pt}}m{#1}}
\usepackage{multirow}

\usepackage{svg}
\usepackage{amsmath}

\begin{document}
\title{Learning for Detecting Norm Violation\\ in Online Communities}

\titlerunning{Norm Violation Detection}

\author{Thiago Freitas dos Santos \and
Nardine Osman \and
Marco Schorlemmer}

\authorrunning{TF dos Santos et al.}

\institute{Artificial Intelligence Research Institute, IIIA-CSIC\\ Barcelona, Catalonia, Spain
}

\maketitle              

\begin{abstract}
In this paper, we focus on normative systems for online communities. The paper addresses the issue that arises when different community members interpret these norms in different ways, possibly leading to unexpected behavior in interactions, usually with norm violations that affect the individual and community experiences. To address this issue, we propose a framework capable of detecting norm violations and providing the violator with information about the features of their action that makes this action violate a norm. We build our framework using Machine Learning, with Logistic Model Trees as the classification algorithm. Since norm violations can be highly contextual, we train our model using data from the Wikipedia online community, namely data on Wikipedia edits. Our work is then evaluated with the Wikipedia use case where we focus on the norm that prohibits vandalism in Wikipedia edits. 

\end{abstract}

\section{Introduction} 
\label{sec:introduction}

The aligned understanding of a norm is an essential process for the interaction between different agents (human or artificial) in normative systems. Mainly because these systems take into consideration norms as the basis to specify and regulate the relevant behavior of the interacting agents \cite{jones1993characterisation}. This is especially important when we consider online communities in which different people with diverse profiles are easily connected with each other. In these cases, misunderstandings about the community norms may lead to interactions being unsuccessful. Thus, the goals of this research are: 1) to investigate the challenges associated with detecting when a norm is being violated by a certain member, usually due to a misunderstanding of the norm; and 2) to inform this member about the features of their action that triggered the violation, allowing the member to change their action to be in accordance with the understanding of the community, thus helping the interactions to keep running smoothly. To tackle these goals, our main contribution is to provide a framework capable of detecting norm violation and informing the violator of why their action triggered a violation detection. 

The proposed framework is using data, that belongs to a specific community, to train a Machine Learning (ML) model that can detect norm violation. We chose this approach based on studies showing that the definition of what is norm violation can be highly contextual, thus it is necessary to consider what a certain community defines as norm violation or expected behavior \cite{al2019detection,chandrasekharan2019crossmod,van2003DeonticLogic}. 

To investigate norm violations, this work is specifically interested in norms that govern online interactions, and we use the Wikipedia community as a testbed, focusing on the article editing actions. This area of research is not only important due to the high volume of interactions that happen on Wikipedia, but also for the proper inclusion and treatment of diverse people in these online interactions. For instance, studies show that, when a system fails to detect norm violations (e.g., hate speech or gender, sexual and racial discrimination), the interactions are damaged, thus impacting the way people interact in the community~\cite{KishonnaDiversity2018,mclean2019female}. 

Previous works have dealt with norms and normative systems, proposing mechanisms for norm conflict detection \cite{NormConflict2018}, norm synthesis \cite{morales2018off}, norm violation on Wikipedia \cite{west2011multilingual,anandDeepLearningViolation2019} and other online communities, such as Stack Overflow \cite{cheriyan2020norm} and Reddit \cite{chandrasekharan2019crossmod}.\removeText{and Twitter \cite{kwokRacism2013}} However, our approach differs mainly in three points: 1) implementing an ML model that allows for the interpretation of the reasons leading to detecting norm violation; 2) incorporating a taxonomy to better explain to the violator which features of their actions triggered the norm violation, based on the results provided by our ML model; and 3) codifying actions in order to represent them through a set of features \change{acquired from previous knowledge about the domain}, which is necessary for the above two points\removeText{(the learning of the ML model and the taxonomy of action features)}. Concerning the last point, we note that our framework does not consider the action as is, but a representation of that action in terms of features and the relation of those features to norm violation \change{(as learned by the applied ML model)}. For the Wikipedia case, we represent the action of editing articles based on the categorization introduced in~\cite{west2011multilingual}, with features such as: the measure of profane words, the measure of pronouns, and the measure of Wiki syntax/markup (the details of these are later presented in Section \ref{subSec:wikipediaTaxonomy}). 


To build our proposed framework, this work investigates the combination of two main algorithms: 1) the Logistic Model Tree, the algorithm responsible for classifying an article edit as a violation or not;\removeText{The dataset used for training is the vandalism on Wikipedia editions corpus} and 2) the K-Means, the clustering algorithm responsible for grouping the features that are most relevant for detecting a violation. The information about the relevant features is then used to navigate the taxonomy and get a simplified taxonomy of these relevant features.

Our experiments describe how the ML model was built based on the training data provided by Wikipedia, and the results of applying this model to the task of vandalism detection in Wikipedia's article edits illustrate how our approach can reach a precision of 78,1\% and a recall of 63,8\%. Besides, the results also show that our framework can provide information about the specific group of features that affect the probability of an action being considered a violation, and we make use of this information to provide feedback to the user on their actions.

The remainder of this paper is divided as follows. Section~\ref{sec:background} presents the basic mechanisms used by our proposed framework. Section~\ref{sec:framework} describes our framework, while Section~\ref{sec:useCase} presents its application to the Wikipedia edits use case, and Section~\ref{sec:experimentsResults} presents our experiment and its results. The related literature is presented in Section~\ref{sec:literatureReview}. We then describe our conclusions and future work in Section~\ref{sec:conclusion}.

\section{Background} 
\label{sec:background}

This section aims to present the base concepts upon which this work is built. We first start with the description of the taxonomy\removeText{(Section~\ref{subSec:taxonomy})}, which we intend to use to formalize a community's knowledge about the features of the actions. Next, we describe the ML algorithms applied to build our framework. First, the Logistic Model Tree (LMT) algorithm\removeText{(Section~\ref{subSec:LMT})}, which is used to build the model responsible for detecting possible vandalism; and second, the K-Means algorithm\removeText{(Section~\ref{subSec:kmeans})}, responsible for grouping the features of the action that are most relevant for detecting violation.

\subsection{Taxonomy for Action Representation}
\label{subSec:taxonomy}

In the context of our work, an action (executed by a user in an online community) is represented by a set of features. Each of these features describes one aspect of the action being executed, i.e., the composing parts of the action. The goal of adopting this approach is to equip our system with an adaptive aspect, since by modelling an action as a set of features allows the system to deal with different kinds of actions (in different domains). For example, we could map the action of participating in an online meeting by features, such as: amount of time present in the meeting; volume of message exchange; and rate of interaction with other participants. Besides, in the context of norm violation, the proposed approach can use these features to explain which aspects of an action were problematic. 

Defining an action through its features gives information about different aspects of the action that might have triggered a violation. However, it is still necessary to find a way to present this information to the violators. The idea is that this information must be provided in a human-readable way, allowing the users to understand what that feature means and how different features are related to each other. With these requirements in mind, we propose the use of a taxonomy to present this data.\removeText{A taxonomy is a way to classify similar concepts into groups, which has been used in different areas, specially in Biology~\cite{kirk96Taxonomy}.} This classification scheme provides relevant information about concepts of a complex domain in a structured way~\cite{kirk96Taxonomy}, thus handling the requirements of our solution. 

We note that, in this work, the focus is not on {\it building} a taxonomy of features. Instead, we assume that the taxonomy is provided with their associated norms. Our system uses this taxonomy, navigating it to select the relevant features. The violator is then informed about the features (presented as a subsection of the larger taxonomy) that triggered the violation \change{detected by our model}.

\subsection{Logistic Model Tree}
\label{subSec:LMT}


With respect to the domain of detecting norm violations in online communities, interpreting the ML model is an important aspect to consider.\removeText{is an important aspect to consider. This need encourages the creation of a solution that offers a tool capable of explaining, to the members of a community, the reasons behind the model's decisions.} Thus, if a community is interested in providing the violator with information about the features of their action that are indicative of violation, then the proposed solution needs to work with a model that can correctly identify these problematic features.

\removeText{In ML, there are different approaches to create a model, }In this work, we are interested in\removeText{the concept of} supervised learning, which is the ML task of finding a map between the input and the output.\removeText{To do that, the model must be trained with labeled data, containing examples of input-output pairs~\cite{russell2002artificial}.} Several algorithms exist that implement the concepts of supervised learning, e.g., artificial neural networks and tree induction. We are most concerned with the ability of these algorithms to generate interpretable outputs, i.e., how the model explains the reasons for taking a certain decision. As such, the algorithm we chose that contains this characteristic is the tree induction algorithm.

The ability to interpret the tree induction model is provided by the way a path is defined in this technique (basically a set of {\em if-then} statements), which allows our model to find patterns in the data, present the path followed by the model and consequently provide the reasons that lead to that conclusion.

Although induction trees have been a popular approach to solve classification problems\removeText{ for a long time\cite{safavian1991survey,ghiasiDT2020}}, this algorithm also presents some disadvantages. This has prompted Landwehr et al.~\cite{landwehr2005logistic} to propose the Logistic Model Tree (LMT) algorithm, which adds logistic regression functions at the leaves of the tree. 

In logistic regression, there are two types of variables: the independent and the dependent variables. The goal is to find a model able\removeText{to map the relationship between the independent and dependent variables, i.e., a function that} to describe the effects of the independent variables on the dependent ones. In our context, the output of the model is\removeText{in the interval [0,1], which is useful to indicate the probability that an event is going to happen, or in our context,} responsible for predicting the probability of an action being classified as norm violation. 

Dealing with odds is an interesting \change{aspect} present in logistic regression, since the increase in a certain variable indicates how the odds changes for the classification output, in this case the odds indicate the effect of the independent variables on the dependent ones. Besides, another important aspect is the equivalence of the natural log of the odds ratio and the linear function of the independent variables, represented by equation \ref{eq:lnOdds}:

\begin{eqnarray}
\label{eq:lnOdds}
	ln (\frac{p}{1-p}) & \gets & \beta_0 + \beta_1 x_1
\end{eqnarray}
where \(ln\) is the logarithm of the odds ratio, \(p\) [0,1] is\removeText{a value between 0 and 1,} the probability of an event occurring. \(\beta\) represents the parameters of the model, in our case the \change{weights for} features of the action. After calculating the natural logarithm, we can then use the inverse of the function to get our estimated regression equation:

\begin{eqnarray}
\label{eq:regressionEstimation}
	\hat p & \gets \frac{\epsilon^{\beta_0 + \beta_1 x_1}}{1+\epsilon^{\beta_0 + \beta_1 x_1}}
\end{eqnarray}
where $\hat p$ is the probability estimated by the regression model. 

With these characteristics of logistic regression, we can see how this technique can be used to highlight attributes (independent variables) that have \change{relevant} influence over the output of the classifier probability. 


Landwehr et al.~\cite{landwehr2005logistic} demonstrate how neither of the two algorithms described above (Tree Induction and Logistic Regression) is better than the other.\removeText{One conclusion presented by~\cite{landwehr2005logistic} is that the performance of the methods depends upon the domain of investigation and the number of training data available. So, to tackle these issues, the LMT algorithm partitions the instance space in regions of constant class with estimations of the probabilities, while having a more gradual change, provided by the estimations from the logistic regression model.} Thus, to tackle the issues present in these two algorithms, LMT adds to the leaves of the tree a logistic regression function. 

\begin{figure}[ht]
    \begin{center}
    \includegraphics[width=0.8\columnwidth]{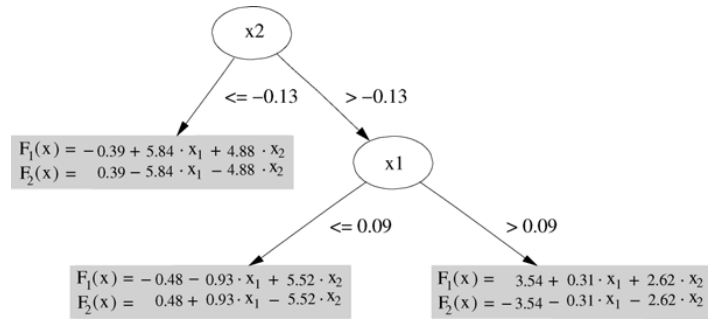}
	\caption{An example of a tree built by the LMT algorithm\cite{landwehr2005logistic}. \change{X\(_1\) and X\(_2\) are features present in the dataset. F\(_1\) and F\(_2\) are the equations found by the logistic regression model, describing the weights for each feature present in the training dataset.}}
	\label{fig:lmtExample}
	\end{center}
\end{figure}

Figure~\ref{fig:lmtExample} presents the description of a tree generated by the LMT algorithm. With a similar process as the standard decision tree, the LMT algorithm obtains a probability estimation as follow: first, the feature is compared to the value associated with that node\removeText{(if the node contains a nominal attribute, with \(k\) values, then this node has \(k\) child nodes. But, if the value is numerical, then the feature is compared to a threshold, going left in the tree if the value is below the threshold and going right otherwise)}. This step is repeated until the algorithm has reached a leaf node, when the exploration is completed. Then the logistic regression function determines the probabilities for the class, as described by equation~\ref{eq:regressionEstimation}.

\subsection{K-Means Clustering Method}
\label{subSec:kmeans}

K-Means is a clustering algorithm with the goal of finding a number \(K\) of clusters in the observed data, attempting to group the most `similar' data points together. This algorithm has been used successfully in different applications, such as\removeText{network intrusion detection~\cite{jianliangIntrusionKMeans2009} and} feature learning~\cite{RubaiatFeaturesKmeans2018} and computer vision~\cite{zheng2018image}. To achieve this goal, K-Means clusters the data using the nearest mean to the cluster center (calculating the squared Euclidean distance), thus reducing the variance within the group~\cite{rai2010survey}.


In this work, the K-Means algorithm can be used to group the features that may indicate an action as violation \change{(we use the features' weights multiplied by their input values as indication of relevance for the classification probability)}. First, after detecting a possible violation, the ML model provides the K-Means algorithm with the set of features present in the logistic regression and their associated \change{values (the input multiplied by the weight)}. Then, based on the values of these features, the algorithm is responsible for separating the features in two groups: 1) those \change{that our model found with highest values, i.g., the most relevant for the vandalism classification}; and 2) those \change{with the lowest values, e.g., less relevant for the vandalism classification}. Lastly, the output of K-Means informs the framework which are the most relevant features for detecting violations (i.e. the first group), which the framework can then use to navigate the taxonomy and present a selected simplified taxonomy of relevant features to the violator.

\section{Framework for Norm Violation Detection (FNVD)}
\label{sec:framework}

This section presents the main contribution of our work, the framework for norm violation detection (FNVD). The goal of this framework is to be deployed in a normative system so that when a violation is detected, the system can enforce the norms by, say, prohibiting the action.   

The main component of our framework is the machine learning (ML) algorithm behind the detection of norm violations, specifically the LMT algorithm of Section~\ref{subSec:LMT}. 
%
%
An important aspect to take into consideration, when using this algorithm, is the data needed to train the model. In \removeText{the context of }our work, the community must provide the definitions \removeText{and constitutions }of norm violations through a dataset that exemplifies actions that were previously labeled as norm violations.\removeText{Different communities use different methods to build this dataset, ranging from using only feedback from the community itself (e.g., from the moderators of the community) to hiring a service to classify the data (such as the Mechanical Turk).} Thus, here we are using data provided by Wikipedia, gathered using Mechanical Turk~\cite{Potthast2010OverviewOT}.



\begin{figure}[ht]
    \begin{center}
        \includegraphics[width=1.0\columnwidth]{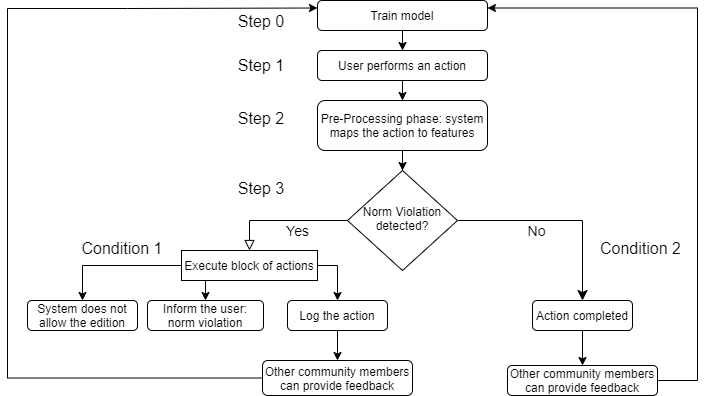}
    	\caption{How the framework works when deployed in an online community.}
    	\label{fig:frameworkDiagram}
	\end{center}
\end{figure}

After defining the data source, our proposed approach essentially 1) collects the data \removeText{that will be }used to train the LMT model; 2) trains the LMT model to detect possible violations and to learn the action's features relevant to norm violations; and 3) when violations are detected, according to the LMT model's results, then the action responsible for the violation is rejected and the violator is informed about the features of their action \change{that triggered the model output}. Furthermore, in both cases (when actions are labelled as violating norms or not), we suggest that the framework collects feedback from the members of the community, which can then be used as new data to retrain the ML model. This is important as we strongly believe that communities and their members evolve, and what may be considered a norm violation today might not be in the future. For example, imagine a norm that states that hate speech is not allowed. Agreeing on the features of hate speech may change from one group of people to another and may also change over time. Consider the evaluation of the N-word, which is usually seen as a serious racial offense and can automatically be considered a text that violates the ``no hate speech" norm. However, imagine a community of African Americans frequently saluting each other with the phrase ``Wussup nigga'' and the ML model classifying their text as hate speech. Clearly, human communities do not always have one clear definition of concepts like hate speech, violation, freedom of speech, etc. The framework, as such, must have a mechanism to adapt to the views of the members of its community, as well as adapt to the views that may change over time. While we leave the adaptation part for future work, we highlight its need in this section, and prepare the framework to deal with such adaptions, as we illustrate in Figure~\ref{fig:frameworkDiagram}.

To further clarify how our framework would act to detect a norm violation when deployed in a community, it is essential to explore the diagram in Figure~\ref{fig:frameworkDiagram}. Step~0 represents the training process of the LMT model, which is a fundamental part of our approach because it is in this moment that the rules for norm violation are specified. Basically, after training the model, our framework would have identified a set of rules that describe norm violation. We can portray these rules as a conjunction of two elements: 1) the tree that is built by the LMT algorithm on top of the collected data; and 2) the weights presented in the leaves of the tree. These weights are the parameters of the estimated regression equation that defines the probability of norm violation (depicted in Equation \ref{eq:regressionEstimation}). With the trained LMT model, the system starts monitoring every new action performed in the community (Step~1). In Step~2, the system maps the action to features that the community defined as descriptive of that action, which triggers the LMT model to start working to detect if that action is violating (or not) any of the norms. Step~3 presents the two different paths that can be executed by our system. If the action is detected as violating a norm (Condition~1), then we argue that the system must execute a sequence of steps to guarantee that the community norms are not violated: 1) the system does not allow the action to persist (i.e., action is not executed); 2) the system presents to the user information about which action features were the most relevant for our model to detect the norm violation, and the taxonomy of the relevant features is presented\removeText{. In the case of our work, these features can be seen as the composing parts of the actions, i.e., we are separating the actions into features (which are used to train our machine learning model, details in Section \ref{subSec:taxonomy})}; 3) the action is logged by the system, allowing other community members to give feedback about the edit attempt, thus providing the possibility of these members flagging the action as a non-violation. The feedback collected from the users can later be used to continuously train (Step~0) our LMT model (future work). However, if the executed action is not detected as violating a norm (Condition~2), then the system can proceed as follows. The action persists in the system (i.e., action is executed), and since any model may incorrectly classify some norm violation as non-violation, the system allows the members of the community to give feedback about that action, providing the possibility of flagging an already accepted action as a violation. Getting people's feedback on violations that go unnoticed by the model is a way to allow the system to adapt to new data (people's feedback) and update \change{the definitions of norm violations} by continuously training the LMT model (Step~0).

To obtain the relevant features for the norm violation classification (Condition~1), we use the K-Means algorithm\removeText{(Section \ref{subSec:kmeans})}. In our context, due to the estimated logistic regression equation, the LMT model provides the weights for each feature \change{multiplied by the value of these features for the action}. \change{This indicates} the influence of the features on the model's output (i.e., the probability of an action being classified as norm violation). With the weights \change{and specific values for the features}, \change{the K-Means algorithm can group the set of features that present the highest multiplied values, which are the ones we assume that contribute the most for the probability of norm violation}. Then, by searching the taxonomy using the group of relevant features, our system can provide the taxonomy structure of the features that trigger norm violation, this is useful due to the explanatory and interpretation characteristics of a taxonomy. The aim of providing this information is to clarify to the member of the community performing the action, what are the problematic aspects of their action \change{as learned by our model}.

\section{The Wikipedia Vandalism Detection Use Case}
\label{sec:useCase}

We focus on the problem of detecting vandalism in Wikipedia article edits. This use case is interesting because Wikipedia is an online community where norms such as `no vandalism' may have different interpretations by different people\removeText{, and norm violations can affect community experience}. In what follows, we first present the use case's domain, followed by the taxonomy used by our system, and finally, an illustration of how our proposed framework may be applied to this use case.

\subsection{Domain}
\label{subSec:domain}

Wikipedia~\cite{wiki}\removeText{\footnote{\url{https://en.wikipedia.org}}} is an online encyclopedia, containing articles about different subjects in any area of knowledge. It is free to read and edit, thus any person with a device connected to the internet can access it and edit\removeText{the content of} its articles. Due to the openness and collaborative structure of Wikipedia, the system is subject to different interpretations of what is the community's expectation concerning how content should be edited. To help address this issue, Wikipedia has compiled a set of rules, the Wikipedia norms~\cite{Potthast2010OverviewOT}, to maintain and organize its content.


Since we are looking for an automated solution for detecting norm violations by applying machine learning mechanisms, the availability of data becomes crucial. Wikipedia provides data on what edits are marked as vandalism, where vandalism annotations were provided by Amazon's Mechanical Turk. Basically, every article edit was evaluated by at least three people (Mechanical Turks) that decided whether the edit violates the `no vandalism' norm or not. In the context of our work, the actions performed by the members of the community are the Wikipedia users' attempts to edit articles, and the norm is {\em ``Do not engage in vandalism behavior"} (which we\removeText{sometimes} refer to as the `no vandalism' norm). It is this precise dataset that we have used to train the model that detects norm violations. We present an example of what is considered a vandalism in a Wikipedia article edit, where a user edited an article by adding the following text: {\em ``Bugger all in the subject of health ect."} 

\subsection{Taxonomy Associated with Wikipedia's `No Vandalism' Norm}
\label{subSec:wikipediaTaxonomy}

An important step in our work is to map actions to features and then specify how they are linked to each other.\removeText{As illustrated earlier, we assume a taxonomy of these features are already provided with the norm. In our Wikipedia use case, however, we had to manually create this taxonomy.} We manually created a taxonomy to describe these features \change{ by separating them in categories that describe their relation with the action.\footnote{For the complete taxonomy, the reader can refer to \url{https://bit.ly/3sQFhQz}}} \change{In this work, we consider the 58 features described in~\cite{west2011multilingual} and 3 more that were available in the provided dataset: LANG\_MARKUP\_IMPACT, the measure of the addition/removal of Wiki syntax/markup; LANG\_EN\_PROFANE\_BIG and LANG\_EN\allowbreak\_PROFANE\_BIG\_IMPACT, the measure of addition/removal of English profane words. In the dataset, features ending with \_IMPACT are normalized by the difference of the article size after edition.} The main objective of this taxonomy is to help our system present to the violator an easy-to-read explanation of the reasons why their article edit was marked as violating a norm by our model, \change{specifying the features with highest influence to trigger this violation.} 

\begin{figure}[h]
    \begin{center}
        \includegraphics[width=0.8\columnwidth]{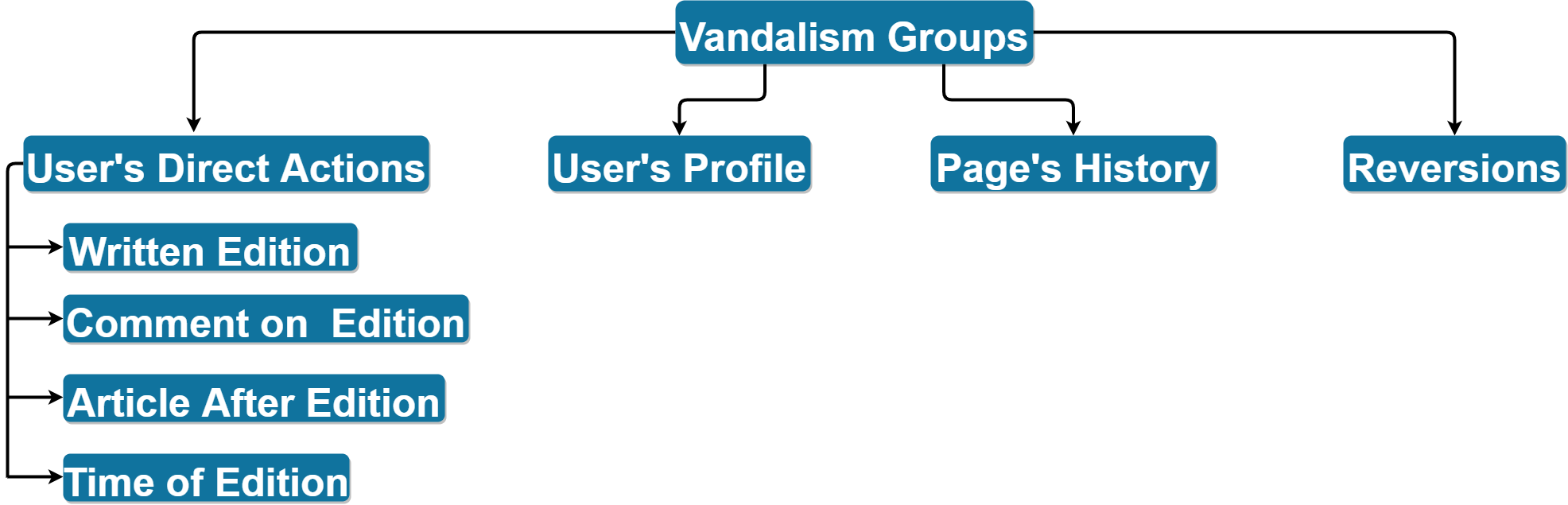}
	    \caption{Taxonomy associated with Wikipedia's `no vandalism' norm.}
	    \label{fig:vandalismDetectionTaxonomy}
    \end{center}
\end{figure}

To further explain our taxonomy approach, we present in Figure~\ref{fig:vandalismDetectionTaxonomy} the constructed taxonomy for Wikipedia's `no vandalism' norm. We observe that features can be divided in four main groups. The first is user's direct actions, which represent aspects of the user's article editing action, e.g., adding a text. This group is further divided in four sub-groups: a) written edition, which contains features about the text itself that is being edited by the user; b) comment on the edition, which contains features about the comments that users have left on that edition; c) article after edition, which contains features about how the edited article changed after the edition was completed; and d) time of edition, which contains features about the time when the user made their edition. The second group is the user's profile, general information about the user. The third is the page's history,\removeText{information about the page that the user is editing, e.g., } how the article changed with past editions. The last group is reversions, which is essentially information on past reversions.\footnote{A reversion is when an article is reverted back to a version before the vandalism occurred.} \change{In total, these groups have 61 features, but due to simplification purpose, Table~\ref{tab:ExampleFeatures} only presents a subset of those features. \removeText{that are part of these groups:}} 

\begin{table}[h]
\scriptsize
	\caption{Example of Features present in the taxonomy groups.}\label{tab:ExampleFeatures}
	\centering
    \begin{tabular}{l|llll}
	    \hline
		\multicolumn{1}{c|}{{Group}} & {{Features}} \\
		\hline
		\textbf{Written Edition} & LANG\_ALL\_ALPHA; LANG\_EN\_PRONOUN\\
		\textbf{Comment on Edition} & COMM\_LEN; COMM\_LEN\_NO\_SECT\\
		\textbf{Article After Edition} & SIZE\_CHANGE\_RESULT; SIZE\_CHANGE\_CHARS\\
		\textbf{Time of Edition} & TIME\_TOD; TIME\_DOW\\
		\textbf{User's Profile} & HIST\_REP\_COUNTRY; USER\_EDITS\_DENSITY\\
		\textbf{Page's History} & PAGE\_AGE; WT\_NO\_DELAY\\
		\textbf{Reversions} & HASH\_REVERTED; HASH\_IP\_REVERT\\
		\hline
    \end{tabular}
\end{table}


\subsection{FNVD Applied to Wikipedia Vandalism Detection}
\label{subSec:frameworkWikipediaVandalism}


It this section, we first describe an example of how our framework can be configured to be deployed in the Wikipedia community. First, the community provides the features and the taxonomy describing that feature space (see Figure~\ref{fig:vandalismDetectionTaxonomy}).\removeText{This information is essential because the idea of the framework is not only to detect that an edition might be a vandalism, but also inform violators about the features that have contributed to the model classifying their edit as vandalism.} Then,\removeText{with this set of features and its taxonomy,} our framework trains the LMT model to classify norm violations based on the data provided (Step~0 of Figure~\ref{fig:frameworkDiagram}), which must contain examples of what that community understands as norm violation and regular behavior. 

In the context of vandalism detection on Wikipedia, the relevant actions performed by the members of the community are the attempts to edit Wikipedia articles. Following the diagram in Figure~\ref{fig:frameworkDiagram}, when a user attempts to edit an article (Step~1), our system will analyze this edit. We note here that our proposed LMT model does not work with the action itself, but the features that describe it. As such, it is necessary to first find the features that represent the performed action\removeText{(the Wikipedia article edit)}. Thus, in Step~2\removeText{of Figure \ref{fig:frameworkDiagram}}, there is a pre-processing phase responsible for mapping actions to the features associated to the norm in question. For example, \change{an article about Asteroid was edited with the addition of the text {\em `` i like man!!"}. After getting this edition text, the system can compute the values (as described in \cite{west2011multilingual}) for the 61 features, which are used to calculate the vandalism probability}. \change{For brevity reasons, we only show the values for some of these features}: 
\begin{enumerate} 
\item LANG\_ALL\_ALPHA, the percentage of text which is alphabetic: 0,615385;
\item WT\_NO\_DELAY, the calculated WikiTrust score: 0,731452;    
\item HIST\_REP\_COUNTRY, measure of how users from the same country as the editor behaved: 0,155146.
\end{enumerate}

\removeText{For that, first we have to go through the path of the tree, to determine the logistic regression equation that we are using to calculate the probability of vandalism, } 

\change{After calculating the values for all features,} the LMT model can evaluate if this article edit is considered `vandalism' or not. In the case of detecting vandalism (Condition~1 of Figure~\ref{fig:frameworkDiagram}), the system does not allow the edition to be recorded on the Wikipedia article, and it presents to the violator two inputs. The first is the set of features of their edit that have the highest influence on the model's decision to detect the vandalism. \change{To get this set, after calculating the probability of vandalism \removeText{with the logistic regression equation }(as depicted in Equation~\ref{eq:regressionEstimation}), the LMT model provides the features that present a positive relationship with the \change{output}. These `positive features' are then used by K-Means to create the group with the most relevant ones (Table~\ref{tab:ExamplePositiveFeatures} presents an example of this process)}. The second input is the selected part of the taxonomy related to chosen set of features, providing further explanation of those features that triggered the norm violation. Additionally, the system will log the attempt to edit the article, which eventually may trigger feedback collection that can at a later stage be used to retrain our model. 

\begin{table}[h]
	\caption{\change{List of features that positively affects the probability of vandalism detection. Total Value is the multiplication between the feature's values and the features' weights\removeText{ (defined by the LMT model)}. The most relevant features, as found by K-Means, are marked with an (*).}}
	\label{tab:ExamplePositiveFeatures}
	\centering 
    \begin{tabular}{l|llll}
	    \hline
		\multicolumn{1}{c|}{{Features}} & {{Total Value}} \\
		\hline
		\textbf{WT\_NO\_DELAY*} & 1.08254896\\
		\textbf{HIST\_REP\_COUNTRY*} & 0.899847\\
		\textbf{LANG\_ALL\_ALPHA*} & 0.7261543\\
		\textbf{HASH\_REC\_DIVERSITY} & 0.15714292\\
		\textbf{WT\_DELAYED} & 0.12748878\\
		\textbf{LANG\_ALL\_CHAR\_REP} & 0.12\\
		\textbf{HIST\_REP\_ARTICLE} & 0.093548\\
		\hline
    \end{tabular}
\end{table}

\begin{figure}[h]
    \begin{center}
        \includegraphics[width=0.6\columnwidth]{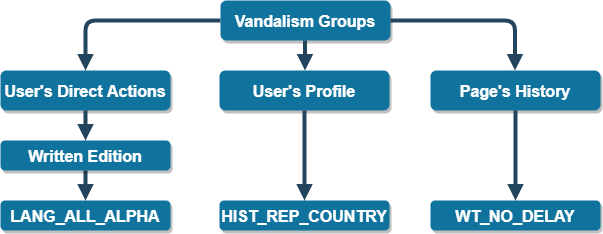}
    	\caption{Taxonomy for part of the features that were \change{most relevant for the vandalism classification. These features are then presented to the user with a descriptive text.}}
    	\label{fig:featuresVandalismTaxonomy}
	\end{center}
\end{figure}

\removeText{Since our proposed approach navigates the taxonomy to get information about the features that are relevant for norm violation detection, an example demonstrating how this taxonomy is used in this case is interesting for clarification purposes. }\change{The features \(WT\_NO\_DELAY\), \(HIST\_REP\_COUNTRY\) and \(LANG\_ALL\allowbreak\_ALPHA\) were indicated by K-Means as the most relevant for the classification of vandalism}. With this information, our framework can search the taxonomy for the relevant features and then automatically retrieve the simplified taxonomy structure for these three specific features, as shown in Figure \ref{fig:featuresVandalismTaxonomy}.
\removeText{Besides, a description text is also presented to the violator, explaining what that feature represents.}

However, in case the system classifies the article edit as `non-vandalism' (Condition~2 of Figure~\ref{fig:frameworkDiagram}), the Wikipedia article is updated according to the user's article edit and community members may provide feedback on this new article edit, which may later be used to retrain our model (as explained in Section~\ref{sec:framework}).


\section{Experiments and Results}
\label{sec:experimentsResults}

The goal of this section is to describe how the proposed approach was applied for detecting norm violation in the domain of Wikipedia article edits, with an initial attempt to improve the interactions in online communities. Then, we demonstrate and discuss the results achieved.

\subsection{Experiments}
\label{subSec:experiments}

\change{Data on vandalism detection in Wikipedia articles~\cite{west2011multilingual} were used for the experiments.}\removeText{Our experiments were conducted with data on vandalism detection in Wikipedia articles~\cite{west2011multilingual}}\removeText{provided by a competition the Wikipedia community organized to evaluate how different models deals with the automatization process of vandalism detection \cite{Potthast2010OverviewOT}} This dataset has 61 features and 32,439 instances for training (with 2,394 examples of vandalism editions and 30,045 examples of regular editions). The model was \change{trained with WEKA~\cite{weka} and} evaluated using 10 folds cross-validation. 

\begin{figure}[ht]
    \begin{center}
        \includegraphics[width=\columnwidth]{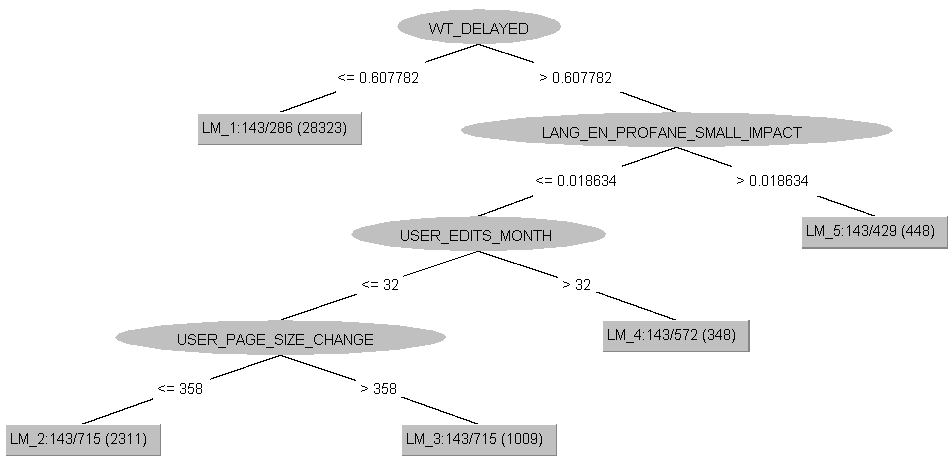}
    	\caption{The built model for the vandalism detection, using Logistic Model Tree.}
    	\label{fig:LMTModelForVandalismDetection}
	\end{center}
\end{figure}

\removeText{We built the model showed in Figure \ref{fig:LMTModelForVandalismDetection}, this was achieved by using the WEKA tool~\cite{weka}.\removeText{\footnote{\url{https://www.cs.waikato.ac.nz/ml/weka/}}} As described in Section \ref{subSec:LMT}, the leaves in the tree represent the estimated logistic equations that will be used to calculate the probability of a certain edition being classified as vandalism. In these equations, we have different \change{weight values} for each feature of our data.\footnote{Trained model available at:\url{https://bit.ly/3gBBkwP}}\removeText{, as shown in the list below:}} 
\removeText{\begin{table}[h]
	\caption{Some of the weights for features in the logistic regression equation.}\label{tab:weightsLMTModel}
	\centering
    \begin{tabular}{l|ll}
	    \hline
		\multicolumn{1}{c|}{Feature} & \multicolumn{1}{c}{Value} \\
		\hline
		LANG\_EN\_PROFANE\_BIG\_IMPACT & 40.33\\
		LANG\_EN\_PRONOUN\_IMPACT & 26.67\\
		LANG\_ALL\_MARKUP\_IMPACT & -29.79\\
		\hline
    \end{tabular}
\end{table}
}
\removeText{
\begin{enumerate} 
\item LANG\_EN\_PROFANE\_BIG\_IMPACT, value: 40,33;
\item LANG\_EN\_PRONOUN\_IMPACT, value: 26,67;
\item LANG\_ALL\_MARKUP\_IMPACT, value: -29,79;
\end{enumerate}
The positive values indicate a positive relationship between the feature and the classification probability, and the negative values otherwise.
}

\subsection{Results}
\label{subSec:results}


The first important information to note is how the LMT model performs when classifying vandalism in Wikipedia editions. In Figure \ref{fig:LMTModelForVandalismDetection}, it is possible to see the model that was built to perform the classification task.\footnote{Trained model available at: \url{https://bit.ly/3gBBkwP}} The tree has four decision nodes and five leaves in total. Since the LMT model uses logistic regression at the leaves, the model has five different estimated logistic regression equations, each of these equations outputs' the probability of an edition being a vandalism. 

\removeText{The LMT model performs well to classify the Wikipedia Vandalism dataset, since }The LMT model correctly classifies 96\% of instances in general. However, when we separate the results in two groups, vandalism editions and regular editions, it is possible to observe a difference in the model's performance. For the\removeText{instances that are } regular editions, the LMT model achieves a precision of 97,2\%, and a recall of 98,6\%. While for\removeText{the instances of} vandalism editions, the performance of the model drops, with a precision of 78,1\% and a recall of 63,8\%. This decrease\removeText{in the percentage for vandalism detection } can be explained by how the dataset was separated and the number of vandalism instances, which consequently leads to an unbalanced dataset. In the dataset, the total number of vandalism instances is 2,394 and the other 30,045 instances are of regular editions. A better balance between the number of vandalism editions and regular edition should improve our classifier, thus in the future we are exploring other \change{model configurations (e.g., ensemble models) to handle data imbalance}. 

\begin{figure}[ht]
    \begin{center}
        \includegraphics[width=0.8\columnwidth]{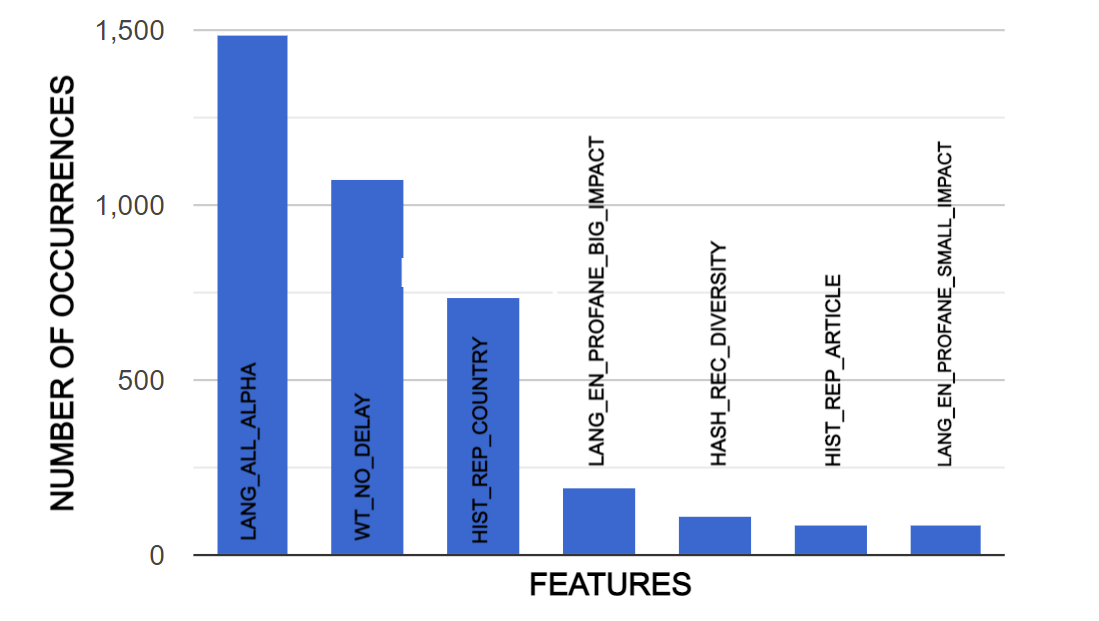}
	    \caption{Number of occurrences of relevant features in vandalism detection.}
	    \label{fig:GraphNumberOfOccurrences}
	\end{center}
\end{figure}

The influence of each feature on determining the probability of a norm violation is provided by the LMT model \change{(as assumed in this work, feature influence is model specific, meaning that a different model can find a different set of relevant features)}.\removeText{In this regard, }\removeText{The proposed framework presents }\removeText{presenting the features that are most relevant when a vandalism is detect by our model.} The graph in Figure~\ref{fig:GraphNumberOfOccurrences} shows the number of times a feature is \change{classified as relevant by the built model}. Some features appear in most of the observations, indicating how important they are to detect\removeText{this kind of} vandalism. Future work shall investigate if this same behavior (some features present in the actions have more influence than other features to define the norm violation probability) can be detected in other domains.

\removeText{An interesting characteristic from the Wikipedia Vandalism Detection data-set is }\removeText{We note that the feature}``LANG\_ALL\_ALPHA" recurrently appears as relevant when vandalism is detected. This happens because this feature presents, as estimated by the LMT model, a positive relationship with the norm violation, meaning that when a vandalism edition is detected, this feature is usually relevant for the classification. 

\removeText{With the presentation of the relevant results, it is also important to demonstrate how a vandalism is detected by our system\removeText{, describing the relevant features that lead to this classification}. The text edition {\em `` They alos went to parties and drank tequila every saturday!”} is a vandalism edition. In our model, the estimated logistic regression equations calculated the relevant features, and the K-Means algorithm grouped these features together: LANG\_ALL\_ALPHA, WT\_NO\_DELAY, and HIST\_REP\_COUNTRY. 
\removeText{
\begin{enumerate} 
\item LANG\_ALL\_ALPHA;
\item WT\_NO\_DELAY; 
\item HIST\_REP\_COUNTRY.
\end{enumerate}
}}

\section{Related Work}
\label{sec:literatureReview}
\done{BEFORE SUBMISSION: fix the right way to cite authors depending upon the manner they are present in the text.}
In this section, we present the most relevant works related to that reported in this paper. \change{Specifically, we reference the relevant literature that uses ML solutions to learn the meaning of a violation, then use that to detect violations in online communities.}\removeText{Specifically, we reference the literature relevant to the area of norm violation detection in online communities, with applications of Machine Learning (ML) solutions.}\removeText{We are interested in any sort of online communities, considering from question-answer websites to social medias.} In addition to the specific works presented below, it is also worth to mention a survey that studies a variety of research in the area, focusing on norm violation detection in the domains of hate speech and cyberbullying \cite{al2019detection}.

Also investigating norm violation in Wikipedia but using the dataset from the comments on talk page edits, Anand and Eswari~\cite{anandDeepLearningViolation2019} present a Deep Learning (DL) approach to classify a comment as abusive or not.\removeText{Besides, the proposed model also further classifies the violation in more specific terms, like toxic, obscene and identity hate.} Although the use of DL is an interesting approach to norm violation detection, we focus on offering interpretability, i.e., providing \change{features our model found as relevant} for the detection of norm violation. While the DL model in~\cite{anandDeepLearningViolation2019} does not provide such information. 

The work by Cherian et al. \cite{cheriyan2020norm} explores norm violation on the Stack Overflow (SO) community. This violation is studied by analyzing the comments posted on the site, which can contain hate speech and abusive language. The authors state that the SO community could become less toxic by identifying and minimizing this kind of behavior, which they separate in two main groups: generic norms and SO specific norms.\removeText{Besides, \cite{cheriyan2020norm} proposes the idea of a recommendation system, responsible for informing the person violating the norm about rephrasing options for the comment being posted.} There are two important similarities between our works: 1) both studies use labeled dataset from the community, considering the relevant context; and 2) the norm violation detection workflow. The main difference is that we focus on the interpretation of the reasons that indicate a norm violation as detected by our model, providing information to the user so they can decide which specific features they are changing. This is possible because we are mapping the actions into features, while Cheriyan at al. \cite{cheriyan2020norm} work directly with the text from the comments, which allows them to focus on providing text alternatives to how the user should write their comment. 

Chandrasekaran et al. \cite{chandrasekharan2019crossmod} build a system for comment moderation in Reddit, named Crossmod.\removeText{The authors work on top of previous work \cite{chandrasekharan2019hybrid}.} Crossmod is described as a sociotechnical moderation system designed using participatory methods (interview with Reddit moderators).\removeText{The system is intended to help moderating comments posted on a Reddit community (subreddit).} To detect norm violation, Crossmod uses a ML back-end, formed by an ensemble of classifiers\removeText{, specifically Crossmod has 108 classifiers}.\removeText{From these, 100 were trained in specific subreddits and 8 trained to detect general norm violation on Reddit.} Since there is an ensemble of classifiers, the ML back-end was trained using the concept of cross-community learning, which uses data from different communities to detect violation in a specific target community.\removeText{\cite{chandrasekharan2019crossmod} assumes that the moderation task depends on context, thus the main goal is not to automatically remove comments that are classified as violation, but rather give this information to moderators so they can decide the action to take.} Like our work, Crossmod uses labeled data from the community to train the classifiers and the norm violation detection workflow follows the same pattern. However, different from our approach, Chandrasekaran et al. \cite{chandrasekharan2019crossmod}  use textual data directly, not mapping to features. Besides, Crossmod do not provide to the user information on the parts of the action that triggered the violation classifier. \removeText{so although they have cross-community learning, the classifiers focus on textual tasks.} \removeText{The focus on working directly with textual data differs. This differs from our approach of using the features of the actions, since not working directly with text gives our framework the flexibility to be deployed in communities with different kind of task. Besides, Crossmod do not provide to the user information on the parts of the action that triggered the violation classifier.} 

Considering another type of ML algorithm, Di Capua et al. \cite{capuaBullying2016} build a solution based on Natural Language Processing (NLP) and Self-Oganizing Map (SOM) to automatically detect bullying behavior on social networks.\removeText{Although the solution was fine-tuned to work on Twitter, the approach was also applied to Youtube and Formspring.} The authors decided to use an unsupervised learning algorithm because they wanted to avoid the manual work of labeling the data, the assumption is that the dataset is huge and by avoiding manual labelling, they would also avoid imposing a priori bias about the possible classes. This differs from our assumptions since we regard the data/feedback from the community as the basis to deal with norm violation.

One interesting aspect about these studies is that they are either in the realm of hate speech or cyberbullying, which can be understood as a sub-group of norm violation by formalizing hate speech and cyberbullying in terms of norms that a community should adhere to. Researchers are interested in these fields mainly due to the damage that violating these norms can cause in the members of an online community, and due to the available data to study these communities. 

\section{Conclusion and Future Work}
\label{sec:conclusion}

The proposed framework, combining machine learning (Logistic Model Trees and K-Means) and taxonomy exploration, is an initial approach on how to detect norm violations. In this paper, we focused on the issue of norm violation assuming violations may occur due to misunderstandings of norms originated by the diverse ways people interpret norms in an online community. To study norm violation, our work used a dataset from Wikipedia's vandalism edition, which contains data about Wikipedia article edits that were considered vandalism.

The framework described in this work is a first step towards detecting vandalism, and it provides relevant information about the problems (features) of the action that led to vandalism. Further investigation is still needed to get a measure of how our system would improve the interactions in an online community. The experiments conducted in our work show that our ML model has a precision of 78,1\% and a recall of 63,8\% when classifying data describing vandalism.

Future work is going to focus on the use of feedback from the community members to continuously train our ML model, as explained in Section~\ref{sec:framework}. The idea is to apply an online training approach to our framework, so when a community behavior changes, that would be taken to indicate a new view on the rules defining the norm, and our ML model should adapt to this new view. 


Throughout this investigation, we have noticed that the literature mostly deals with norm violation that focus either on hate speech or cyberbullying. We aim that our approach can be applied to other domains (not only textual), thus we are planning to explore domains with different actions to analyze how our framework deals with a different context (since these domains would have a different set of actions to be executed in an online community).

\done{
This an idea for future work, but I don't think it fits this article now.

Another future direct that we plan to follow is to provide the problematic editions with personalized suggestion. Considering that we have access to profile information, we can then group the editions by certain profile groups, thus offering suggestions to different groups, rather than to the whole community. This is particularly interesting because the users may act in different ways and perhaps this way of behavior is not shared in a community level, only in a more profile specific level. Thus, when this group is interacting, the norm can be flexible, not being enforced as the general norm of the community.}

\section*{Acknowledgements}
This research has received funding from the European Union's Horizon 2020 FET Proactive project ``WeNet – The Internet of us'', grant agreement No 823783, as well as the RecerCaixa 2017 funded ``AppPhil'' project.

\bibliographystyle{splncs04}
\bibliography{references.bib}

\begin{thebibliography}{10}
\providecommand{\url}[1]{\texttt{#1}}
\providecommand{\urlprefix}{URL }
\providecommand{\doi}[1]{https://doi.org/#1}

\bibitem{NormConflict2018}
{Aires}, J.P., {Monteiro}, J., {Granada}, R., {Meneguzzi}, F.: Norm {C}onflict
  {I}dentification using {V}ector {S}pace {O}ffsets. In: IJCNN. pp.~1--8 (2018)

\bibitem{al2019detection}
Al-Hassan, A., Al-Dossari, H.: Detection of {H}ate {S}peech in {S}ocial
  {N}etworks: a {S}urvey on {M}ultilingual {C}orpus. In: 6th International
  Conference on Computer Science and Information Technology. pp. 83--100 (2019)

\bibitem{anandDeepLearningViolation2019}
{Anand}, M., {Eswari}, R.: Classification of {A}busive {C}omments in {S}ocial
  {M}edia using {D}eep {L}earning. In: 2019 3rd International Conference on
  Computing Methodologies and Communication (ICCMC). pp. 974--977 (2019)

\bibitem{chandrasekharan2019crossmod}
Chandrasekharan, E., Gandhi, C., Mustelier, M.W., Gilbert, E.: Crossmod: {A}
  {C}ross-{C}ommunity {L}earning-{B}ased {S}ystem to {A}ssist {R}eddit
  {M}oderators. Proc. ACM Hum.-Comput. Interact.  \textbf{3}(CSCW) (Nov 2019)

\bibitem{cheriyan2020norm}
Cheriyan, J., Savarimuthu, B.T.R., Cranefield, S.: Norm {V}iolation in {O}nline
  {C}ommunities--{A} {S}tudy of {S}tack {O}verflow {c}omments. arXiv preprint
  arXiv:2004.05589  (2020)

\bibitem{capuaBullying2016}
{Di Capua}, M., {Di Nardo}, E., {Petrosino}, A.: Unsupervised {C}yber
  {B}ullying {D}etection in {S}ocial {N}etworks. In: ICPR. pp. 432--437 (2016)

\bibitem{kirk96Taxonomy}
Fiedler, K.D., Grover, V., Teng, J.T.: An {E}mpirically {D}erived {T}axonomy of
  {I}nformation {T}echnology {S}tructure and {I}ts {R}elationship to
  {O}rganizational {S}tructure. Journal of Management Information Systems
  \textbf{13}(1),  9--34 (1996)

\bibitem{KishonnaDiversity2018}
Gray, K.L.: Gaming out {o}nline: {B}lack {L}esbian {I}dentity {D}evelopment and
  {C}ommunity {B}uilding in {X}box {L}ive. Journal of Lesbian Studies
  \textbf{22}(3),  282--296 (2018)

\bibitem{jones1993characterisation}
Jones, A.J., Sergot, M., et~al.: On the {C}haracterisation of {L}aw and
  {C}omputer {S}ystems: {T}he {N}ormative {S}ystems {P}erspective. Deontic
  logic in computer science: normative system specification pp. 275--307 (1993)

\bibitem{landwehr2005logistic}
Landwehr, N., Hall, M., Frank, E.: Logistic {M}odel {T}rees. Machine learning
  \textbf{59}(1-2),  161--205 (2005)

\bibitem{mclean2019female}
McLean, L., Griffiths, M.D.: Female {G}amers' {E}xperience of {O}nline
  {H}arassment and {S}ocial {S}upport in {O}nline {G}aming: {A} {Q}ualitative
  {S}tudy. International journal of mental health and addiction
  \textbf{17}(4),  970--994 (2019)

\bibitem{morales2018off}
Morales, J., Wooldridge, M., Rodr{\'\i}guez-Aguilar, J.A.,
  L{\'o}pez-S{\'a}nchez, M.: Off-{L}ine {S}ynthesis of {E}volutionarily
  {S}table {N}ormative {S}ystems. Autonomous agents and multi-agent systems
  \textbf{32}(5),  635--671 (2018)

\bibitem{Potthast2010OverviewOT}
Potthast, M., Holfeld, T.: Overview of the 1st {I}nternational {C}ompetition on
  {W}ikipedia {V}andalism {D}etection. In: CLEF (2010)

\bibitem{rai2010survey}
Rai, P., Singh, S.: A {S}urvey of {C}lustering {T}echniques. International
  Journal of Computer Applications  \textbf{7}(12), ~1--5 (2010)

\bibitem{RubaiatFeaturesKmeans2018}
{Rubaiat}, S.Y., {Rahman}, M.M., {Hasan}, M.K.: Important {F}eature {S}election
  {A}ccuracy {C}omparisons of {D}ifferent {M}achine {L}earning {M}odels for
  {E}arly {D}iabetes {D}etection. In: International Conference on Innovation in
  Engineering and Technology. pp.~1--6 (2018)

\bibitem{van2003DeonticLogic}
van~der Torre, L.: Contextual {D}eontic {L}ogic: {N}ormative {A}gents,
  {V}iolations and {I}ndependence. Annals of mathematics and artificial
  intelligence  \textbf{37}(1-2),  33--63 (2003)

\bibitem{west2011multilingual}
West, A.G., Lee, I.: Multilingual vandalism detection using
  language-independent \& ex post facto evidence. In: CLEF Notebooks (2011)

\bibitem{wiki}
{Wikipedia contributors}: Wikipedia --- {W}ikipedia{,} the free encyclopedia
  (2021), \url{\url{https://en.wikipedia.org/wiki/Wikipedia}}, [Online;
  accessed 22-Feb-2021]

\bibitem{weka}
Witten, I.H., Frank, E., Hall, M.A.: The {WEKA} {W}orkbench. In: Data Mining:
  Practical Machine Learning Tools and Techniques, pp. 403--406. Morgan
  Kaufmann (2011)

\bibitem{zheng2018image}
Zheng, X., Lei, Q., Yao, R., Gong, Y., Yin, Q.: Image {S}egmentation {B}ased on
  {A}daptive {K}-{M}eans {A}lgorithm. EURASIP J Image Video Process
  \textbf{2018}(1),  1--10 (2018)

\end{thebibliography}
\end{document}